\documentclass[aip,jap,reprint,floatfix,citeautoscript,longbibliography,superscriptaddress]{revtex4-2}
\usepackage{graphicx,bm}
\usepackage{amsmath,amssymb,bbold}
\usepackage[usenames]{color}
\usepackage{hyperref}
\usepackage[capitalise]{cleveref}
\hypersetup{colorlinks=true, linkcolor=blue, citecolor=blue, urlcolor=blue}
\begin{document}

\title{Antimagnonics}

\author{J. S. Harms}
\affiliation{Institute for Theoretical Physics, Utrecht University, 3584CC Utrecht, The Netherlands}

\author{H. Y. Yuan}
\affiliation{Institute for Theoretical Physics, Utrecht University, 3584CC Utrecht, The Netherlands}

\author{Rembert A. Duine}
\affiliation{Institute for Theoretical Physics, Utrecht University, 3584CC Utrecht, The Netherlands}
\affiliation{Department of Applied Physics, Eindhoven University of Technology, P.O. Box 513, 5600 MB Eindhoven, The Netherlands}

\date{\today}

\begin{abstract}
Magnons are the quanta of collective spin excitations in magnetically-ordered systems and manipulation of magnons for computing and information processing has witnessed the development of ``magnonics". A magnon corresponds to an excitation of the magnetic system from its ground state and the creation of a magnon thus increases the total energy of the system. In this perspective, we introduce the antiparticle of a magnon, dubbed the antimagnon, as an excitation that lowers the magnetic energy. We investigate the stability and thermal occupation of antimagnons and verify our theory by micromagnetic simulations. Furthermore, we show how the concept of antimagnons yields a unified picture to understand the magnonic analog of the Klein effect, magnonic black-hole horizons, and magnonic black-hole lasing. Our work may stimulate fundamental interest in antimagnons, as well as their applications to spintronic devices.
\end{abstract}


\maketitle

\section{Introduction}
Magnons are quasi-particle excitations of spins in ordered magnets, and the manipulation of magnons for information processing has spurred on magnon spintronics or magnonics \cite{ChumakNP2015}. As information carriers, magnons can propagate even in magnetic insulators, which, in principle, avoids the heating problems accompanying a charge current in the traditional transistor technologies. In the last decade, the generation, manipulation and read-out of magnon currents have attracted significant attention. Recently, the scope of magnonics has been extended to the quantum regime, where magnon quantum states and their integration with well known quantum platforms including qubits, cavity photons and phonons are investigated \cite{YuanPR2022}.

As opposed to magnons, their antiparticles called antimagnons are rarely studied. In general, the excitation of magnons from the ground state of a magnet increases the energy of the magnetic system and thus magnons carry positive energy. Here, we define antimagnons as quasi-particle excitations that lower the energy of the system. This definition is a bit more restrictive than the most general definition of antimagnons, namely, to define the antimagnon as carrying spin opposite to magnons, irrespective of the energy of the antimagnon. Our definition is motivated by the finding that interesting physics arises precisely in the case that antimagnons carry negative energy. This is because in this situation, they can be coupled to magnons, and be used to, for example, amplify the magnons. (Note that our definition of antimagnons is different from the ``antimagnon" used to discuss the motion of the N\'{e}el vector in the absence of magnetization oscillations in a specific class of ferrimagnets driven by electric fields \cite{Turov2007}.)

Following our definition, a dilemma immediately arises: how could a physical system be stable while supporting negative-energy excitations? Here, we answer this question by considering excitations on top of energetically unstable states that are dynamically stabilized by effectively reversing the sign of the damping. Excitations on top of an energetically unstable state, such as a magnet pointing against its effective field, by definition lower the energy of the system, and are thus negative-energy excitations. Normally, such an energetic instability is in the presence of dissipation accompanied by a dynamical instability that allows the energetic instability to unfold, most likely resulting in dynamics whereby the magnetization is reversed. In the absence of dissipation, or when the effective sign of the dissipation is reversed by some form of external driving, the energetically-unstable state is stabilized and yields stable negative-energy excitations.
Indeed, this concept was introduced previously by us to stabilize antimagnons in spintronics, where the gain may come from spin-orbital torque \cite{Joren2021}, spin-transfer torque \cite{roldan2017magnonic,harms2021dynamically}, optical driving and other mechanisms that may reverse the sign of the effective Gilbert damping \cite{CaoPRB2022}.

Following the introduction of antimagnons, phenomena including the magnonic black hole, magnonic black-hole lasing, and magnonic Klein effects are understood in a unified picture. Namely, these latter examples rely on magnonic versions of particle-antiparticle generation that arises from coupling of magnons to antimagnons. The introduction of antimagnons not only provides a solid-state platform to study high-energy physics, but may benefit the amplification of magnon currents in spintronic devices due to magnon-antimagnon interaction. In this perspective, we first give an elaborate introduction to antimagnons. Hereafter, we briefly discuss how their introduction leads to the magnon analog of the Klein paradox, magnonic lasing and black-hole horizons. We end with a conclusion and outlook.

\section{Formalism} \label{sec:formalism}

In the sections below, we formally introduce antimagnons. While some of this discussion is well-known in the field of analogue gravity~\cite{faccio2013analogue,barcelo2011analogue,novello2002artificial}, it may not be familiar to researchers working in magnetism, spintronics, and magnonics. Because of this, we try to be detailed and complete in what follows. For readers who wish to skip these details, we give here a brief summary: We discuss how the linearized Landau-Lifshitz-Gilbert (LLG) equation yields an eigenvalue problem that is well-known in the Bogoliubov theory of excitations in superfluids, and which leads to the introduction of a specific conserved norm. This eigenvalue problem yields pairs of eigenfrequencies of which the eigenmodes have opposite norm and which physically correspond to the same excitation. Because of this doubling, it is sufficient to consider only positive or negative frequencies.

For linearization around the true magnetic ground state --- referred to as the energetically-stable situations --- the positive-norm modes have positive frequencies, while the negative-norm modes have negative frequencies. Upon quantization, the former correspond to magnons, while the latter would correspond to antimagnons. Typically, one restricts oneself to positive frequencies and therefore antimagnons do not need to be introduced when discussing excitations over the true ground state.

The situation changes if one considers excitations on top of a state that is not the magnetic ground state, which we refer to as the energetically unstable situation. In this case, there may be positive-norm states with negative frequency, and negative-norm states with positive frequency. Restricting oneself again to positive frequencies, one now has to consider the negative-norm states with positive frequencies. We define these excitations to be antimagnons, as they carry opposite spin to magnons and physically correspond to negative-energy excitations. This is because the system is now linearized around a metastable state, and the excitation lowers the total energy.

Adding any amount of dissipation would normally make the energetically-unstable situation dynamically unstable as well: because the environment is able to dissipate energy, the system dynamically evolves to its true ground state. The energetically-unstable situation may be made dynamically stable by external pumping that effectively reverses the sign of the damping. Below we discuss in detail the example of how this may be achieved by spin-orbit torque.    Once the energetically-unstable state is dynamically stabilized it yields stable antimagnon excitations that may be coupled to magnons. Examples of the physics that results from this coupling are discussed in Sec.~\ref{sec:apps}

\subsection{Spin waves}\label{sec:classical-spin-waves}

\begin{figure}
	\centering
	\includegraphics[width=\columnwidth]{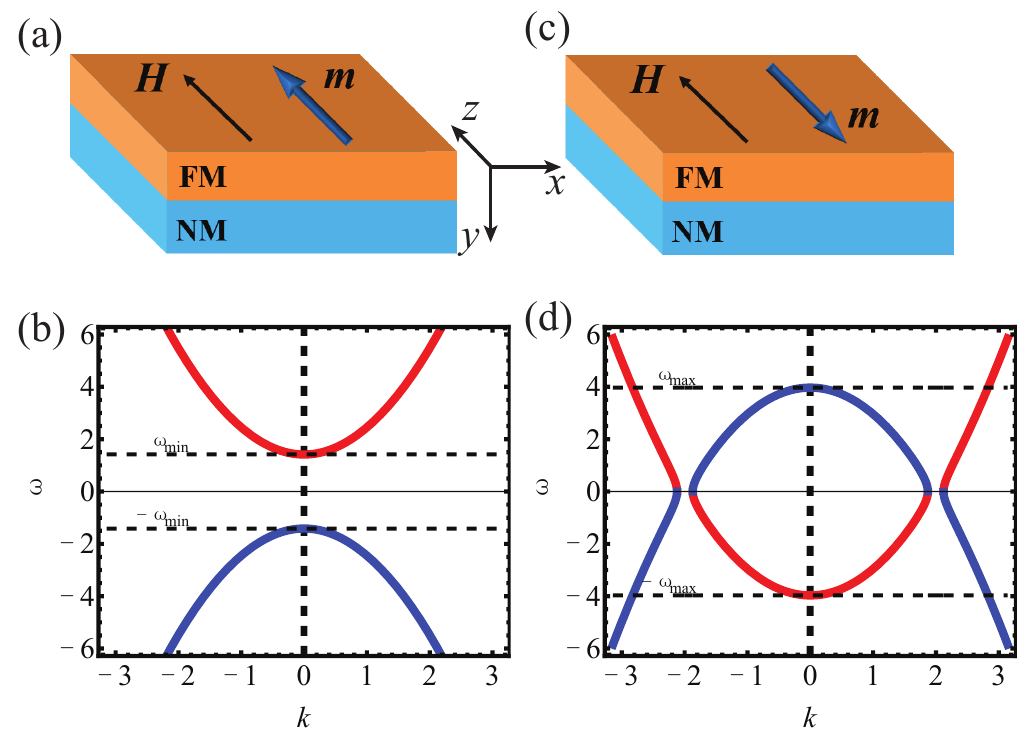}\\
	\caption{ Scheme of a magnetic thin film adjacent to a heavy metal layer with magnetization parallel and antiparallel to the external field. The bottom panel shows the corresponding dispersion relations of magnetic excitations.}\label{magnon_antimagnon_setup}
\end{figure}

To be concrete, we consider an insulating ferromagnetic (FM) thin film with hard-axis anisotropy in the $y$-direction adjacent to a heavy-metal layer (HM), {as shown in Fig. \ref{magnon_antimagnon_setup}}. Furthermore, we consider an external magnetic field in the $ z $ -direction. The discussion below may be easily generalized to other examples of magnetic anisotropies and fields.
Far below the Curie temperature, the dynamics is well described by the LLG equation for the direction $\mathbf{n}$ of the magnetization with spin-orbit torques (SOT),
\begin{equation}\label{LLG}
	\frac{\partial \mathbf{n}}{\partial t}=-\gamma \mathbf{n} \times \mathbf{h}_{\mathrm{eff}} + \alpha \mathbf{n} \times \frac{\partial \mathbf{n}}{\partial t} + J_s \mathbf{n} \times \hat{z} \times \mathbf{n},
\end{equation}
with $\gamma$  the gyromagnetic ratio, $\alpha$ the Gilbert damping and $ J_s $ the strength of SOTs generated by the spin current which depends on the current flowing in the HM layer, the spin Hall angle of the HM and the properties of the interface.
The above equation describes damped precession around the effective magnetic field $\mathbf{h}_{\mathrm{eff}}=-\delta E/(M_s\delta\mathbf{n}) $, with $M_s$ the saturation magnetization and
\begin{equation}\label{eq:effective_hamiltoniaan}
	\begin{aligned}
		E=&\int dV \bigg\{	
		A(\nabla_i\mathbf{n})^2
		-\mu_0H_{\mathrm{e} }M_s n_{z}+\frac{1}{2}Kn_{y}^2
		\bigg\}
	\end{aligned}
\end{equation}
the energy functional of this set-up.
In the above $ A $ is the exchange stiffness,
$ H_{\mathrm{e}} $ is the external magnetic field strength, $\mu_0 $ is the vacuum permeability with $ K $ the anisotropy, which may be caused by magnetocrystalline anisotropy or dipolar interactions.


We introduce spin waves as linearized dynamical fluctuations on top of the static magnetization.
As discussed, we consider the static magnetization to be either in the direction of the external magnetic field or opposite to it, i.e.
$\mathbf{n}_{0}=\pm\hat{z}$. The former ($\mathbf{n}_{0}=\hat{z}$) corresponds to the energetically stable configuration whereas the latter  ($\mathbf{n}_{0}=-\hat{z}$) corresponds to the energetically unstable configuration.
We proceed by introducing the complex field\\
$
\Psi=(1/\sqrt{2})\left(\hat{e}_1+\mathrm{i}\hat{e}_2\right)\cdot\mathbf{n},
$
with $ \hat{e}_1\times\hat{e}_2=\mathbf{n}_0 $.
For convenience we consider $ \hat{e}_1=\hat{x} $, which automatically gives $ \hat{e}_2=\mp\hat{y} $.
When linearizing the LLG equation~\eqref{LLG} in $\delta n_1, \delta n_2$, according to $\mathbf{n} = \mathbf{n}_0 + (\delta n_1, \delta n_2, 0)^T$, it is recast as a Bogoliubov-de-Gennes like equation
\begin{align}\label{eq:linearized_LLG_equation}
	\frac{\mathrm{i}(\mathbb{1}+\mathrm{i}\alpha\sigma_z)\partial_t}{\gamma\mu_0M_s}
	\begin{pmatrix}
		\Psi\\~\,\Psi^*
	\end{pmatrix}
	=
	\left(\mathcal{L}_\pm+\mathrm{i}I_s\mathbb{1}\right)
	\begin{pmatrix}
		\Psi\\~\,\Psi^*
	\end{pmatrix},
\end{align}
with
\begin{align}\label{eq:linearized-LLG-L-matrix}
	\mathcal{L}_\pm
	=
	\left(\Delta\pm h-\Lambda^2\nabla^2\right)\sigma_z+\mathrm{i}\Delta\sigma_y.
\end{align}
In the above, $ \sigma_{y,z} $ are the Pauli matrices and $ \mathbb{1} $ is the $ 2\times2 $ identity matrix, $ \Lambda=\sqrt{2A/\mu_0M_s^2} $ the exchange length,
$ h=H_{\mathrm{e}}/M_s $ the dimensionless external magnetic field,
$ \Delta=K/2\mu_0M_s^2=1/2 $ the dimensionless anisotropy constant
and $ I_s=J_s/\gamma\mu_0M_s $ the dimensionless SOT.

We write the complex wavefunction $ \Psi $ in terms of Bogoliubov modes
\begin{align}\label{eq:bogoliubov_ansatz}
	\Psi(\mathbf{x},t)&=u(\mathbf{x})e^{-\mathrm{i}\lambda t}+v^*(\mathbf{x})e^{\mathrm{i}\lambda^*t}
\end{align}
and for later convenience, we define the dimensionless frequency $ \omega=\lambda/\gamma\mu_0M_s $.
At this point we make several observation.
First of all, the dissipationless limit of~\cref{eq:linearized_LLG_equation} ($\alpha=J_s=0$) is pseudo-Hermitian, in other words
\begin{equation}
	\sigma_z\mathcal{L}_\pm^\dagger\sigma_z=\mathcal{L}_\pm.
\end{equation}
A consequence of this statement is that the inner product $ \langle\Psi,\Psi'\rangle\equiv\langle\Psi|\sigma_z|\Psi'\rangle
=\int dV\left[u^*(\mathbf{x})u'(\mathbf{x})-v^*(\mathbf{x})v'(\mathbf{x})\right] $ and hence the non-positive definite magnon norm
\begin{equation}\label{eq:magnon-norm}
	\|\Psi\|
	=\langle\Psi|\sigma_z|\Psi\rangle
	=\int dV\left(|u|^2-|v|^2\right),
\end{equation}
is conserved in the dissipationless limit.
Secondly,~\cref{eq:linearized_LLG_equation} has the additional symmetry
\begin{equation}
	\sigma_x(\mathcal{L}_\pm+\mathrm{i}\alpha\omega^*\sigma_z+\mathrm{i}I_\mu\mathbb{1})\sigma_x
	=
	-(\mathcal{L}_\pm-\mathrm{i}\alpha\omega\sigma_z+\mathrm{i}I_\mu\mathbb{1})^*.
\end{equation}
This implies that if $ \omega $ is an eigenfrequency of~\cref{eq:linearized_LLG_equation} with eigenmode
$
\begin{pmatrix}
	u&v
\end{pmatrix}^\mathrm{T}
$,
then
$ -\omega^* $ is an eigenfrequency with eigenmode
$
\begin{pmatrix}
	v^*&u^*
\end{pmatrix}^\mathrm{T}
$.
These two modes have opposite norm by construction.
Hence, the two branches of~\cref{eq:linearized_LLG_equation} are related to each other via particle-hole symmetry. In our magnetic system, this doubling is not physical, but merely a result of our choice to describe spin waves using complex scalar fields.
Hence, in order to determine the full dynamics of the system it is sufficient to consider $ \omega>0 $ and take into account the norm of different modes.
As we see later, the sign of the norm in~\cref{eq:magnon-norm} describes whether we work with magnon or antimagnon commutation relations.
\subsubsection{Spin-wave excitations on top of the ground state}
The ground state corresponds to $ \mathbf{n}_0=\hat{z} $.
Because of translation invariance, we introduce spin-wave modes as the Fourier modes of $ u(\mathbf{x}) $ and $ v(\mathbf{x}) $ in the linearized LLG equation~\eqref{LLG}--around the $ +\hat{z} $ direction.
Up to first order in dissipative terms $ \alpha $ and $ I_s $, these spin wave solutions have the following dispersion relation
\begin{equation}\label{eq:spinwave_dispersion}
	\omega_{\mathbf{k}}
	\simeq
	\omega_{\mathbf{k}}^0
	-\mathrm{i}\left(\alpha\left[\Delta+ h+\Lambda^2k^2\right]-I_s\right),
\end{equation}
with $ \omega=\lambda/\gamma\mu_0M_s $ the dimensionless frequency and
\begin{equation}\label{eq:spin-wave-dispersion-ground-state}
	\omega_{\mathbf{k}}^0
	=
	\|\Psi_k\|\sqrt{\left(\Delta+ h+\Lambda^2k^2\right)^2 - \Delta^2},
\end{equation}
the real part of the dispersion relation in which $ \|\Psi_k\|=\pm1 $ is the norm of the modes.
In~\cref{magnon_antimagnon_setup}\textcolor{blue}{(b)} we show the real part of the dispersion relation, where the red curve corresponds to the positive norm mode and the blue curve corresponds to the negative norm mode.
From the stability requirement we discussed in this section we find that the ground state is stable if $ I_s/\alpha<[\Delta+h] $.
Hence, the ground state becomes unstable if it is driven sufficiently strong such that the SOT overcompensates the damping.
This is well known to happen in spin-torque-oscillators.

The energy functional~\eqref{eq:effective_hamiltoniaan} for spin-wave excitations per definition gives us
\begin{align}
	E\equiv\frac{1}{2}\langle\Psi|\sigma_z\mathcal{L}_+|\Psi\rangle.
\end{align}
By expanding $ |\Psi\rangle=\sum_ka_k|\Psi_k\rangle $ in spin-wave eigenmodes, we find
\begin{align}
	E&=
	\frac{1}{2}\sum_{l,k}a_l^*a_k\langle\Psi_l|\sigma_z\mathcal{L}_+|\Psi_k\rangle
	\\\nonumber
	&=
	\frac{1}{2}\sum_{l,k}a_l^*a_k\langle\Psi_l|\sigma_z|\Psi_k\rangle\omega_k^0
	\\\nonumber
	&=
	\frac{1}{2}\sum_{k}|a_k|^2\|\Psi_k\|\omega_k^0,
\end{align}
Thus, the contribution of the spin-wave modes to the energy is given by $ \|\Psi_k\|\omega_k^0 $. Hence, one may choose to work with the positive-norm modes which have positive frequency and the excitation of which leads to an increase of energy. Alternatively, one may work with negative-energy modes with negative frequency, the excitation of which also lead to an increase in energy. The former choice is the conventional one and, upon quantization, leads to conventional magnons. Choosing to restrict oneself to negative frequency would lead to the same magnonic excitations after a particle-hole transformation, but is an unnecessary complication. This, however, changes when one considers spin-wave excitations on top of a metastable state.

\subsubsection{Spin-wave excitations on the metastable state}\label{eq:spin-waves-meta-stable-gound-state}
Now, we consider spin waves on top of the metastable configuration in which the static magnetization is pointing opposite to the external magnetic field, i.e. $ \mathbf{n}_0=-\hat{z} $.
Due to this different quantization axis we find that the complex field becomes
$ \Psi=(1/\sqrt{2})(\hat{x}+\mathrm{i}\hat{y})\cdot\mathbf{n} $ instead of $ \Psi=(1/\sqrt{2})(\hat{x}-\mathrm{i}\hat{y})\cdot\mathbf{n} $
which we used in the previous section.
This, as we will see in~\cref{sec:quantization}, precisely corresponds to the choice of a different-norm branch and thus commutation relation for the excitations after quantization.
As before, spin waves are introduced as the Fourier modes of $ u(\mathbf{x}) $ and $ v(\mathbf{x}) $ in the linearized LLG equation~\eqref{LLG}. We find that the dispersion relation, up to first order in dissipative constants $ \alpha $ and $ I_s $ is given by
\begin{equation}\label{eq:spinwave_dispersion}
	\omega_{\mathbf{k}}
	\simeq
	\omega_{\mathbf{k}}^0
	-\mathrm{i}\left(\alpha\left[\Delta- h+\Lambda^2k^2\right]-I_s\right),
\end{equation}
with
\begin{align}\label{eq:dispersion-relation-meta-stable-equilibrium}
	\mathrm{Re}\left(\omega_{\mathbf{k}}^0\right)
	&=
	\|\Psi_k\|\mathrm{sgn}(\Lambda^2k^2-h)\sqrt{\left(\Delta-h+\Lambda^2k^2\right)^2 - \Delta^2}
\end{align}
the real part of the dispersion relation in which
\[
\|\Psi_k\|
=
\left\{
\begin{array}{r r}
	\pm1,&|\Delta-h+\Lambda^2k^2| > \Delta,\\
	0,&|\Delta-h+\Lambda^2k^2| < \Delta,
\end{array}
\right.
\]
is the norm of the modes.
The real part of the dispersion relation in~\cref{eq:dispersion-relation-meta-stable-equilibrium} is shown in~\cref{magnon_antimagnon_setup}\textcolor{blue}{(d)}, in which the red curve corresponds to the positive norm mode and the blue curve to the negative norm mode.
Similar to the previous subsection, the classical stability follows from the sign of the imaginary part of the spin wave dispersion relation. Here, we see that this configuration is stable once
\begin{equation}\label{eq:stabilty-sot}
	-I_s/\alpha\gtrsim\max(h-\Delta,\Delta/\alpha).
\end{equation}
Thus the magnetization can in principle be held pointing opposite to the external field if the angular momentum injected by the SOT is sufficiently large.
To determine this stability condition we furthermore used
\[
\mathrm{Im}\left(\omega_{\mathbf{k}}^0\right)
=
\left(1-\|\Psi_k\|^2\right)\sqrt{\left(\Delta-h+\Lambda^2k^2\right)^2 - \Delta^2}.
\]
We thus see that reducing the anisotropy $ \Delta $, which yields elliptical magnetization precession, will greatly reduce the critical current needed to keep the metastable state stable.

To continue, the energy functional for the excitations is given by
\begin{align}
	E\equiv\frac{1}{2}\langle\Psi|\sigma_z\mathcal{L}_-|\Psi\rangle.
\end{align}
By expanding in eigenmodes  $ |\Psi\rangle=\sum_ka_k|\Psi_k\rangle $ we find
\begin{align}
	E&=
	\frac{1}{2}\sum_{l,k}a_l^*a_k\langle\Psi_l|\sigma_z\mathcal{L}_-|\Psi_k\rangle
	\\\nonumber
	&=
	\frac{1}{2}\sum_{k}|a_k|^2\|\Psi_k\|\omega_k^0.
\end{align}
Once again we may chose to work with either positive or negative-norm excitations of which the excitation energy is given by $ \|\Psi_k\|\omega_k^0 $ which becomes negative in a specific region of phase space. What is, however, special in this case is that at positive frequencies there now exist both positive-norm excitations and negative-norm excitations. Because they both have positive frequency, they may couple to each other and examples of this coupling are discussed in Sec.~\ref{sec:apps}. Restricting oneself to positive frequencies, one now necesarrily has to explicitly consider the negative-norm modes. Upon quantization, we define these latter modes to be antimagnons.

\subsection{Quantization} \label{sec:quantization}
The formalism discussed in this section is restricted to the dissipationless limit, i.e. $ \alpha\rightarrow0 $ and $ I_s\rightarrow0 $, thereby assuming that the quantization procedure is still valid when turning on the small dissipation.
We start out this section by canonically quantizing the complex scalar field in~\cref{eq:linearized_LLG_equation}.
In this case, canonical quantization should give us
\begin{equation}\label{eq:canonical-quantization}
	[\Psi(\mathbf{x},t),\Psi^\dagger(\mathbf{x}',t)]=\delta(\mathbf{x-x'}),
\end{equation}
since $ \mathrm{i}\Psi^\dagger $ is the canonical momentum associated with $ \Psi $.
For completeness, the canonical momentum of $ \Psi^\dagger $ is $ -\mathrm{i}\Psi $ making the above definition self consistent.
We have seen in~\cref{sec:classical-spin-waves} that the definition of $ \Psi $ depends on the choice of direction of the static magnetization $\mathbf{n}_0$, where a sign change of $\mathbf{n}_0$ implies a complex conjugation of the complex scalar fields, i.e. $ \mathbf{n}_0=\hat{z}\rightarrow-\hat{z}\implies\Psi\rightarrow\Psi^* $.
Hence, if the quantization axis describes the vacuum opposite to the fixed point direction used for linearization one should work with the anomalous commutation relations
\begin{equation}\label{eq:anomalous-canonical-quantization}
	[\Psi(\mathbf{x},t),\Psi^\dagger(\mathbf{x}',t)]=-\delta(\mathbf{x-x'}).
\end{equation}
%

\subsubsection{Wavevector representation}\label{sec:wavevector-representation}
Let us in first instance restrict ourselves to the commutation relation in~\cref{eq:canonical-quantization}.
In order to diagonalize the equation of motion~\eqref{eq:linearized_LLG_equation} we take the following Bogoliubov ansatz
\begin{subequations}\label{eq:psi-in-creation-and-annihilation}
	\begin{align}
	\Psi(\mathbf{x},t)
	=&
	\sum_{k} \left(u_k(x)a_ke^{-\mathrm{i}\omega_kt}+v_k^*(x)a^\dagger_ke^{\mathrm{i}\omega_kt}\right),
	\\
	\Psi^\dagger(\mathbf{x},t)
	=&
	\sum_{k} \left(v_k(x)a_ke^{-\mathrm{i}\omega_kt}+u_k^*(x)a^\dagger_ke^{\mathrm{i}\omega_kt}\right),
\end{align}
\end{subequations}
where $ u_k $ and $ v_k $ are solutions of~\cref{eq:linearized_LLG_equation} and $ a^\dagger $ and $ a $ are the (anti)magnon creation and annihilation operators.
We find that~\cref{eq:canonical-quantization,eq:psi-in-creation-and-annihilation} imply the commutation relations
\begin{equation}\label{eq:commutation-relation-equal-norm}
	 \|\Psi_k\|\|\Psi_{k'}\|[a_k,a^\dagger_{k'}]=\langle\Psi_k,\Psi_{k'}\rangle=\|\Psi_k\|\delta_{k,k'},
\end{equation}
where the last equality in the above equation is only true if both states have the same norm and is otherwise zero.
Thus, if we work with the positive-norm branch we use the commutation relations $ [a_k,a^\dagger_{k'}]=\delta_{k,k'} $, while in the negative norm branch we have to use the anomalous commutations relations $ [a_k,a^\dagger_{k'}]=-\delta_{k,k'} $.
Furthermore, if two operators correspond to states with different norm we get the commutation relation
\begin{subequations}\label{eq:commutation-relations-different-norms}
\begin{align}
	\|\Psi_k\|\|\Psi_{k'}\|[a_k,a_{k'}]
	&=
	\|\Psi_{k'}\|\delta_{k,k'},
	\\
	\|\Psi_k\|\|\Psi_{k'}\|[a^\dagger_k,a^\dagger_{k'}]
	&=
	\|\Psi_{k}\|\delta_{k,k'},
\end{align}
\end{subequations}
and otherwise zero.
The latter is a consequence of the particle-hole symmetry in the equations of motion~\cref{eq:linearized_LLG_equation} for the fields.
From this point onward we relabel creation and annihilation operators of the negative norm branch as $ b^\dagger_k$ and $ b_k $ such that $ [b_k,b^\dagger_{k'}]=-\delta_{k,k'} $ and the operators in the positive as $ a_k $ and $ a^\dagger_k $ with $ [a_k,a^\dagger_{k'}]=\delta_{k,k'} $.
Additionally we relabel solutions of the negative norm branch of~\cref{eq:linearized_LLG_equation} as $ \begin{pmatrix}\tilde{u}&\tilde{v}\end{pmatrix}^T $.
From~\cref{eq:commutation-relation-equal-norm,eq:commutation-relations-different-norms} we unsurprisingly find that the operators in the different branches are related by particle hole symmetry $ b^\dagger_k=a_k $ and $ b_k=a^\dagger_k $.
We find that the field $ \Psi $ may now be written as
\begin{align}\label{eq:canonicalquantization-complex-fields}
	&\begin{pmatrix}
		\Psi(\mathbf{x},t)\\
		\Psi^\dagger(\mathbf{x},t)
	\end{pmatrix}
	=
	\\\nonumber
	&~~~\,\frac{1}{\sqrt{2}}~\,\sum_{\|\Psi_k\|=1} \left[
	\begin{pmatrix}
		u_k\\v_k
	\end{pmatrix}
	a_ke^{-\mathrm{i}\omega_kt+\mathrm{i}kx}
	+
	\begin{pmatrix}
		v^*_k\\u^*_k
	\end{pmatrix}
	a^\dagger_ke^{\mathrm{i}\omega_kt-\mathrm{i}kx}
	\right]
	\\\nonumber
	&+\frac{1}{\sqrt{2}}\sum_{\|\Psi_k\|=-1} \left[
	\begin{pmatrix}
		\tilde{u}_k\\\tilde{v}_k
	\end{pmatrix}
	b_ke^{-\mathrm{i}\omega_kt-\mathrm{i}kx}
	+
	\begin{pmatrix}
		\tilde{v}^*_k\\\tilde{u}^*_k
	\end{pmatrix}
	b^\dagger_ke^{\mathrm{i}\omega_kt+\mathrm{i}kx}
	\right]
\end{align}
The fields $ b^\dagger\equiv a $ and $ b\equiv a^\dagger $ give a doubling of the modes, since the same information is essentially described by the fields $ a $ and $ a^\dagger $ of the positive norm branch.
This can be made more intuitive when we remember that particle hole symmetry gives
$
\begin{pmatrix}
	v^*&u^*
\end{pmatrix}^\mathrm{T}
$
as an eigenmode of~\cref{eq:linearized_LLG_equation} with frequency $ -\omega^* $ if
$
\begin{pmatrix}
	u&v
\end{pmatrix}^\mathrm{T}
$
is an eigenmode with frequency $ \omega $.
We see that the negative norm sector can be mapped onto the positive norm sector via a particle-hole transformation.
Hence, one may chose to work in whatever branch we find convenient. As argued previously, for excitations on top of a metastable state and after restricting oneself to positive frequency, one necessarily has to consider the negative-norm excitations. For the field in Eq.~(\ref{eq:canonicalquantization-complex-fields}), the implies considering both the magnon operators $ a $ and $ a^\dagger $ and the antimagnon operators  $ b$ and $ b^\dagger $.

Let us consider the Hamiltonian for (anti)magnon excitations.
Similar to the previous section this is given by
\begin{align}
	E\equiv\frac{1}{2}\int dV
	\begin{pmatrix}
		\Psi^\dagger&\Psi
	\end{pmatrix}
	\left(\sigma_z\mathcal{L}_\pm\right)
	\begin{pmatrix}
		\Psi\\~\,\Psi^\dagger
	\end{pmatrix}.
\end{align}
Using~\cref{eq:canonicalquantization-complex-fields} we find that the Hamiltonian in second quantized form becomes
\begin{align}
	E&=
	\frac{1}{4}\sum_{\|\Psi_{k,l}\|=1}\left(a_l^\dagger a_k+a_ka^\dagger_l\right)\langle\Psi_l,\Psi_k\rangle\omega_k^0
	\\\nonumber
	&+
	\frac{1}{4}\sum_{\|\Psi_{k,l}\|=-1}\left(b_l b^\dagger_k+b^\dagger_kb_l\right)\langle\Psi_l,\Psi_k\rangle\omega_k^0.
\end{align}
Thus we find
\begin{align}
	E=E_0
	&+\frac{1}{2}\sum_{\|\Psi_{k}\|=1}~\;\|\Psi_k\|\omega_k^0a_k^\dagger a_k
	\\\nonumber&
	+
	\frac{1}{2}\sum_{\|\Psi_{k}\|=-1}\|\Psi_k\|\omega_k^0b_kb^\dagger_k.
\end{align}
In the above $ E_0 $ is the energy without excitations and $ b,b^\dagger $ and $ a,a^\dagger $ the (anti)magon annihilation and creation operators.
As in~\cref{eq:spin-waves-meta-stable-gound-state} we notice that on top of metastable static states there exist negative energy modes which of which, by definition, the product of norm and frequency is negative. This implies that these modes, which we call antimagnons, have opposite handedness with respect to their magnonic counterpart. More discussion on this can be found in~\cref{sec:frequency-representation}.
\subsubsection{Holstein-Primakoff transformation}
In this section we make the connection between the magnons in our canonically quantized theory and the magnons following from the Holstein-Primakoff transformation in a spin model \cite{HP1958}.
We start from the spin Hamiltonian
\begin{equation}\label{HeisenbergHam}
	\mathcal{H}=-J\sum_{\langle ij \rangle}\mathbf{S}_i \cdot \mathbf{S}_j - H\sum_i S_i^z
	+{K_{d}}\sum_i(S_i^y)^2
\end{equation}
where the first term is Heisenberg exchange interaction between neighboring spins with $J$ being exchange coefficient, which relates to the exchange coefficient in~\cref{eq:effective_hamiltoniaan} via $ A=JS^2a^{2-d}/\hbar\gamma $.
The second term is Zeeman energy of spins under an external field $H$ with $H>0$, which is related to $ H_e $ via $ \mu_sM_sH_e=HSa^{-d}/\hbar\gamma $.
Where $ K_d $ is the shape anisotropy which is related to the anisotropy in~\cref{eq:effective_hamiltoniaan} by $ K=K_dS^2a^{-d}/\hbar\gamma $.
The spin operator obeys the commutation relation $[S_i,S_j]=i\epsilon_{ijk}S_k$ with $\epsilon_{ijk}$ the Levi-Civita symbol.
We consider the static state of the system around which we quantize to be $\mathbf{S}=\pm Se_z$. In the true ground state ($\mathbf{S}=+ Se_z$) the decrease of $S_z$ by $\hbar$ corresponds to a magnon excitation.
On the other, on the meta-stable state ($\mathbf{S}=- Se_z$) an increase of $ \hbar $ gives rise to either an magnonic or and anti-magnonic excitation.
The difference between the two is essentially their handedness and their energy. The antimagnonic excitations carry negative energy en have opposite handedness with respect to the magnonic ones.
Formally, we introduce the (anti)magnonic excitations on top of the (meta-)stable ground state via the Holstein-Primakoff transformation
\begin{subequations}\label{magnonHP}
	\begin{align}
		S_i^\pm&=\sqrt{2S-\Psi_i^\dagger \Psi_i} \Psi_i,\\
		S_i^\mp&=\Psi_i^\dagger \sqrt{2S-\Psi_i^\dagger \Psi_i},\\
		S_{iz}&=\pm (S - \Psi_i^\dagger \Psi_i),
	\end{align}
\end{subequations}
such that the commutation relations are given by
\begin{equation}
	[\Psi_i, \Psi_j^\dagger ] = \delta_{ij},~[\Psi_i,\Psi_j]=0.
\end{equation}
We would like to stress here, that the definition of $ S^\pm $ is with respect to the quantization axis $ \pm\hat{z} $.
If $ S\gg \langle\Psi^\dagger_i\Psi_i\rangle $ the spin Hamiltonian can be expanded up to second order in $ \Psi $ and $ \Psi^\dagger $ giving
\begin{align}\label{mangonHamiltonian}
	\mathcal{H}
	=
	&\pm H\sum_i \Psi_i^\dagger \Psi_i
	-JS\sum_{\langle ij \rangle}
	\bigg[
	\Psi_i \Psi_j^\dagger + \Psi_i^\dagger \Psi_j
	-2\Psi_i^\dagger \Psi_i
	\bigg]
	\\\nonumber
	-&(K_dS/2)\sum_i
	\left[\Psi_i^\dagger\Psi_i^\dagger+\Psi_i\Psi_i-2\Psi_i^\dagger\Psi_i\right].
\end{align}
When taking the continuum limit of the above we find
\begin{align}
	\mathcal{H}=\frac{1}{2}\int dV
	\begin{pmatrix}
		\Psi^\dagger&\Psi
	\end{pmatrix}
	\left(\sigma_z\mathcal{L}_\pm\right)
	\begin{pmatrix}
		\Psi\\~\,\Psi^\dagger
	\end{pmatrix},
\end{align}
with $ \mathcal{L}_\pm $ defined in~\cref{eq:linearized-LLG-L-matrix} and
\begin{equation}
	[\Psi(x), \Psi^\dagger(x') ] \rightarrow \delta(x-x'),~[\Psi(x),\Psi(x')]=0.
\end{equation}
We thus find that according to the Holstein-Primakoff transformation our canonically quantized fields $ \Psi $ and $ \Psi^\dagger $ correspond at quadratic order to the spin lowering and raising operators $ S^\pm $ and $ S^\mp $.
Hence, the anomalous operators $ b^\dagger $ and $ b $ describe a vacuum opposite of their quantization axis, up to second order in their fields.


\subsubsection{Frequency representation}\label{sec:frequency-representation}
For processes in which the frequency $ \omega $ is conserved but wavevector is no longer conserved, such as scattering, it is useful to change variables from $ k\rightarrow\omega $.
This representation is useful when coupling magonic excitations to antimagnon excitations, of which examples are given in~\cref{sec:apps}.
Furthermore, this representation makes the need to consider antimagnons on top of the meta-stable state explicit.
We proceed by expanding the $ \Psi $ and $ \Psi^\dagger $ in~\cref{sec:wavevector-representation} in the frequency representation.
Due to doubling of the modes we consider $ \omega>0 $ throughout this section.
Here, we consider $ k $ to be in the continuum. This implies
$ \sum_k\rightarrow\int d^dk $,
furthermore we define $ \|\Psi_k\| $ by
$ \langle\Psi_k|\sigma_z|\Psi_{k'}\rangle=\|\Psi_k\|\delta^d(k-k') $
and $ \|\Psi_k\|\|\Psi_{k'}\|[a_k,a^\dagger_{k'}]=\langle\Psi_k,\Psi_{k'}\rangle\rightarrow\|\Psi_k\|\delta^d(k-k'). $
We will see that a change in coordinates $ k\rightarrow\omega $ proceeds very differently in the case where the static magnetization is pointing in the direction of the external magnetic field, i.e. $ \mathbf{n}_0=\hat{z} $, as compared to the case in which it points against the external magnetic field, i.e. $ \mathbf{n}_0=-\hat{z} $.
Let us in first instance restrict ourselves to the case in which the magnetization is in the direction of the external magnetic field, i.e. $\mathbf{n}_0=\hat{z}$, corresponding to true equilibrium.
From here we perform a coordinate transformation on~\cref{eq:canonicalquantization-complex-fields} to frequency space giving us
\begin{subequations}
	\begin{align}\label{eq:frequency-representation-complex-scalar-field}
	\Psi(\mathbf{x},t)
	=&
	\int_{S^{d-1}} d\vec{\Omega}
	\\\nonumber
	\int_{\omega_\mathrm{min}}^{\infty} d\omega
	&\left[
	u_{\omega,\vec{\Omega}}(\mathbf{x})
	a_{\omega,\vec{\Omega}}\,e^{-\mathrm{i}\omega t}
	+
	v^*_{\omega,\vec{\Omega}}(\mathbf{x})
	a^\dagger_{\omega,\vec{\Omega}}\,e^{\mathrm{i}\omega t}
	\right],
	\\
	\Psi^\dagger(\mathbf{x},t)
	=&
	\int_{S^{d-1}} d\vec{\Omega}
	\\\nonumber
	\int_{\omega_\mathrm{min}}^{\infty} d\omega
	&\left[
	v_{\omega,\vec{\Omega}}(\mathbf{x})
	a_{\omega,\vec{\Omega}}\,e^{-\mathrm{i}\omega t}
	+
	u^*_{\omega,\vec{\Omega}}(\mathbf{x})
	a^\dagger_{\omega,\vec{\Omega}}\,e^{\mathrm{i}\omega t}
	\right].
\end{align}
\end{subequations}
with $ \begin{pmatrix}u_k&v_k\end{pmatrix}^T $ in the positive norm branch, $ \omega_\mathrm{min}=\sqrt{(\Delta+h)^2-\Delta^2} $ the ferromagnetic resonance and the rescaled fields and operators are given by
\begin{align}\label{eq:field-transformation-frequency-representation}
	\begin{pmatrix}
		u_{\omega,\vec{\Omega}}(\mathbf{x})\\
		v_{\omega,\vec{\Omega}}(\mathbf{x})\\
		a_{\omega,\vec{\Omega}}\\
		a^\dagger_{\omega,\vec{\Omega}}
	\end{pmatrix}
	&=
	\sqrt{k^{d-1}}\sqrt{\frac{dk}{d\omega}}
	\begin{pmatrix}
		u_{\vec{k}}(\mathbf{x})\\
		v_{\vec{k}}(\mathbf{x})\\
		a_{\vec{k}}\\
		a^\dagger_{\vec{k}}
	\end{pmatrix}.
\end{align}
These are chosen such that their commutation relations become
	\begin{equation} \left[a_{\omega,\vec{\Omega}},a^\dagger_{\omega',\vec{\Omega}'}\right]
	=
	\delta(\omega-\omega')\delta^{d-1}(\vec\Omega-\vec\Omega').
\end{equation}
Due to the isotropic dispersion in~\cref{eq:spin-wave-dispersion-ground-state,eq:dispersion-relation-meta-stable-equilibrium} the fields in~\cref{eq:field-transformation-frequency-representation} may be simplified further.
Since the dispersion relation is independent of the direction of $ \vec{k} $, the magnitude of $ k $ only depends on the magnitude of $ \omega $.
We may thus express the fields as
$
\begin{pmatrix}
	u_{\omega,\vec{\Omega}}(\mathbf{x})&
	v_{\omega,\vec{\Omega}}(\mathbf{x})
\end{pmatrix}
=
\begin{pmatrix}
	u_\omega&v_\omega
\end{pmatrix}
\exp({\mathrm{i}k(\omega)\vec{\Omega}\cdot\mathbf{x}}),
$
with
$
\begin{pmatrix}
	u_\omega&v_\omega
\end{pmatrix}
=
\sqrt{k^{d-1}}\sqrt{dk/d\omega}
\begin{pmatrix}
	u_k&v_k
\end{pmatrix}
$
in which
$
\begin{pmatrix}
	u_k&v_k
\end{pmatrix}
$
is an eigenvector of~\cref{eq:linearized_LLG_equation} with $ |u_k|^2-|v_k|^2=(2\pi)^{-d} $.

Next, we consider the case in which the equilibrium magnetization is pointing against the external magnetic field, hence $ \mathbf{n}_0=-\hat{z} $.
In this instance the dispersion relation~\cref{eq:dispersion-relation-meta-stable-equilibrium} becomes negative in the positive norm branch and positive for the negative norm branch.
For frequencies below $ \omega_\mathrm{max}=\sqrt{(\Delta-h)^2-\Delta^2}$ we find an additional negative energy mode in every propagation direction.
We stress that these modes carry negative energy since the product of their norm with their frequency is negative, and, following our definition, we refer to these modes the antimagnons.
For $ \omega>\omega_\mathrm{max} $ on the other hand, these antimagnon modes do not exist.
For excitations on top of the meta-stable state the frequency representation of~\cref{eq:canonicalquantization-complex-fields} becomes
\begin{align}\label{eq:frequency-representation-complex-scalar-field-meta-stable-state}
	\Psi(\mathbf{x},t)
	=&
	\int_{S^{d-1}} d\vec{\Omega}
	\\\nonumber
	\int_{0}^{\omega_\mathrm{max}} d\omega
	&\Big[
	u_{\omega,\vec{\Omega}}(\mathbf{x})
	a_{\omega,\vec{\Omega}}\,e^{-\mathrm{i}\omega t}
	+
	v^*_{\omega,\vec{\Omega}}(\mathbf{x})
	a^\dagger_{\omega,\vec{\Omega}}\,e^{\mathrm{i}\omega t}
	\\\nonumber
	+&\ \;
	\tilde{u}_{\omega,\vec{\Omega}}(\mathbf{x})
	b_{\omega,\vec{\Omega}}\,e^{-\mathrm{i}\omega t}
	+
	\tilde{v}^*_{\omega,\vec{\Omega}}(\mathbf{x})
	b^\dagger_{\omega,\vec{\Omega}}\,e^{\mathrm{i}\omega t}
	\Big]
	\\\nonumber
	+\int_{\omega_\mathrm{max}}^{\infty} d\omega
	&\Big[
	u_{\omega,\vec{\Omega}}(\mathbf{x})
	a_{\omega,\vec{\Omega}}\,e^{-\mathrm{i}\omega t}
	+
	v^*_{\omega,\vec{\Omega}}(\mathbf{x})
	a^\dagger_{\omega,\vec{\Omega}}\,e^{\mathrm{i}\omega t}
	\Big],
\end{align}
where the conjugated expression becomes
\begin{align}\label{eq:frequency-representation-complex-scalar-field-meta-stable-state-conjugate}
	\Psi^\dagger(\mathbf{x},t)
	=&
	\int_{S^{d-1}} d\vec{\Omega}
	\\\nonumber
	\int_{0}^{\omega_\mathrm{max}} d\omega
	&\Big[
	v_{\omega,\vec{\Omega}}(\mathbf{x})
	a_{\omega,\vec{\Omega}}\,e^{-\mathrm{i}\omega t}
	+
	u^*_{\omega,\vec{\Omega}}(\mathbf{x})
	a^\dagger_{\omega,\vec{\Omega}}\,e^{\mathrm{i}\omega t}
	\\\nonumber
	&\ \;
	\tilde{v}_{\omega,\vec{\Omega}}(\mathbf{x})
	b_{\omega,\vec{\Omega}}\,e^{-\mathrm{i}\omega t}
	+
	\tilde{u}^*_{\omega,\vec{\Omega}}(\mathbf{x})
	b^\dagger_{\omega,\vec{\Omega}}\,e^{\mathrm{i}\omega t}
	\Big]
	\\\nonumber
	+\int_{\omega_\mathrm{max}}^{\infty} d\omega
	&\Big[
	v_{\omega,\vec{\Omega}}(\mathbf{x})
	a_{\omega,\vec{\Omega}}\,e^{-\mathrm{i}\omega t}
	+
	u^*_{\omega,\vec{\Omega}}(\mathbf{x})
	a^\dagger_{\omega,\vec{\Omega}}\,e^{\mathrm{i}\omega t}
	\Big].
\end{align}
In the above the modes $ \begin{pmatrix}\tilde{u}&\tilde{v}\end{pmatrix} $ have negative norm and positive frequency $ \omega $.
Here, the rescaled fields and operator of the negative norm branch are defined by
\begin{align}\label{eq:field-transformation-frequency-representation-negative-norm}
	\begin{pmatrix}
		\tilde{u}_{\omega,\vec{\Omega}}(\mathbf{x})\\
		\tilde{v}_{\omega,\vec{\Omega}}(\mathbf{x})\\
		b^\dagger_{\omega,\vec{\Omega}}\\
		b_{\omega,\vec{\Omega}}
	\end{pmatrix}
	&=
	\sqrt{k^{d-1}}\sqrt{\frac{dk}{d\omega}}
	\begin{pmatrix}
		\tilde{u}_{\vec{k}}(\mathbf{x})\\
		\tilde{v}_{\vec{k}}(\mathbf{x})\\
		b^\dagger_{\vec{k}}\\
		b_{\vec{k}}
	\end{pmatrix}.
\end{align}
Via the particle-hole transformation $ b_{\omega,\vec{\Omega}}\equiv a^\dagger_{-\omega,\vec{\Omega}} $ and $ b^\dagger_{\omega,\vec{\Omega}}\equiv a_{-\omega,\vec{\Omega}} $  the same mode is described by positive norm and negative frequency $ -\omega $.

In this representation the Hamiltonian becomes
\begin{align}\label{eq:hamiltonian-frequency-representation}
	\mathcal{H}
	=&
	\int_{S^{d-1}}d\vec{\Omega}
	\int_0^{\omega_\mathrm{max}} d\omega
	\omega
	\Big[
	a^\dagger_{\omega,\vec{\Omega}}a_{\omega,\vec{\Omega}}
	-b_{\omega,\vec{\Omega}}b^\dagger_{\omega,\vec{\Omega}}
	\Big]
	\\\nonumber
	+&
	\int_{S^{d-1}}d\vec{\Omega}
	\int_{\omega_\mathrm{max}}^\infty~\,d\omega
	\omega
	\Big[
	a^\dagger_{\omega,\vec{\Omega}}a_{\omega,\vec{\Omega}}
	\Big].
\end{align}
Hence the antimagnons $ b_{\omega,\vec{\Omega}},~b^\dagger_{\omega,\vec{\Omega}} $ or $ a^\dagger_{-\omega,\vec{\Omega}},~a_{-\omega,\vec{\Omega}} $ after a particle-hole transformation, describe excitations with energy $ -\omega $ and hence have opposite handedness as compared to the magnonic excitations.
From the expression in~\cref{eq:frequency-representation-complex-scalar-field-meta-stable-state,eq:frequency-representation-complex-scalar-field-meta-stable-state-conjugate,eq:hamiltonian-frequency-representation} it is clear that for frequencies between $ 0 $ and $ \omega_\mathrm{max} $ one has to consider both magnons and antimagnons, and, as a results, these excitations may couple.

\section{Stability and thermal fluctuations}
In this section we first consider the antimagnonic excitations within micromagnetic simulations. As these incorporate non-linear effects, this numerical study goes beyond the linearized analysis of the previous subsections. Next, we consider the stability of the metastable static magnetization configuration against small transverse fields. As we will see, this yields Cherenkov-like radiation. Finally, we discuss thermal fluctuations  of the metastable state of the magnetization.

\subsection{Numerical verification}
\begin{figure}
  \centering
  \includegraphics[width=0.49\textwidth]{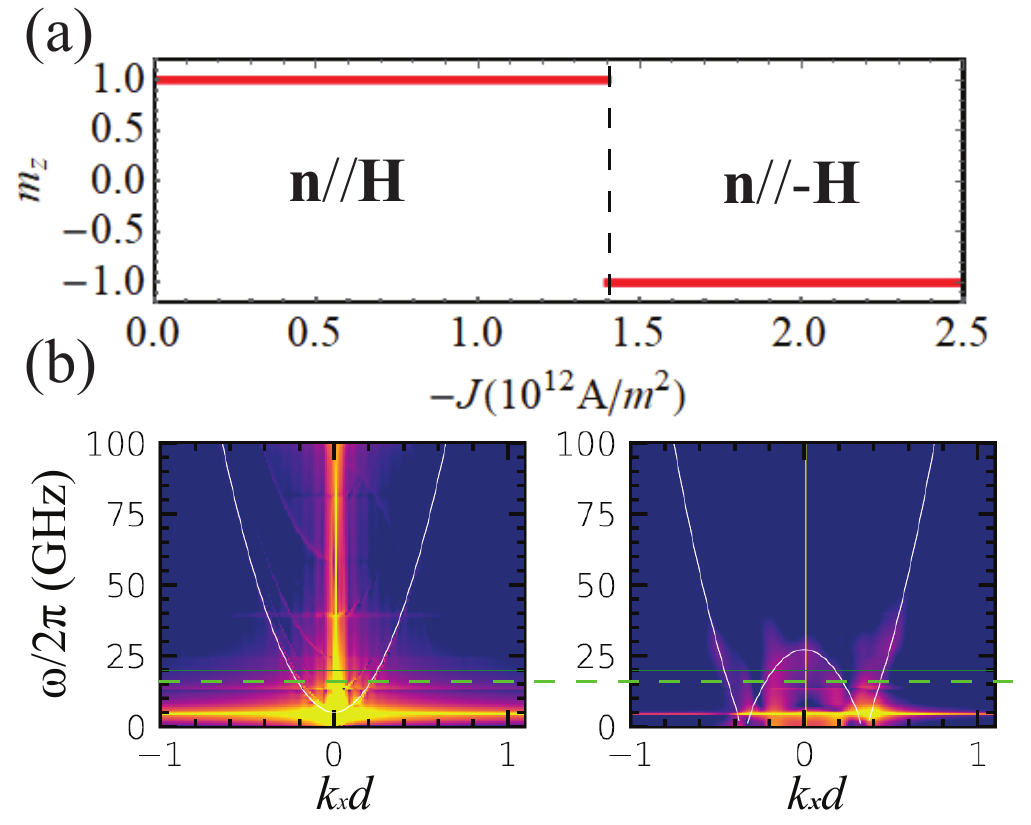}\\
  \caption{(a) Steady magnetization as a function of the applied current density. (b) Magnon (antimagnon) spectrum when the current density is smaller(larger) than the threshold $1.4\times 10^{12}~\mathrm{A/m^2}$. The white lines are analytical calculations of the dispersion relation.}\label{simulated_antimagnon}
\end{figure}

To verify the existence of antimagnon excitations above a dynamically-stable energetically unstable state, we perform micromagnetic simulations on the magnetic layered system shown in Fig~\ref{simulated_antimagnon}. The geometric dimensions of the magnetic film are length $2048$ nm, width $64$ nm and thickness $2$ nm. The strength of SOT depends on the current density as $J_\mathrm{SOT}= J\hbar \theta_{SH}/(2M_s|e|d)$, where $J$ is the current density, $\theta_{SH}$ is the spin-Hall angle of heavy-metal layer, $M_s$ is saturation magnetization, $e$ is electron charge and $d$ is thickness of the magnetic film. The magnetic parameters of a YIG/Pt bilayer are used, i.e., exchange coefficient $A=3.1 \times 10^{-12} ~\mathrm{J/m}$, saturation magnetization $M_s = 1.92 \times 10^5 ~\mathrm{A/m}$, spin-Hall angle $\theta_{SH}=0.1$ \cite{WangPRL2014}, Gilbert damping $\alpha = 10^{-2}$. The magnetic film is discretized into a set of cuboid meshes with dimensions $2 \times 2 \times 2~\mathrm{nm}^3$. The $Mumax^3$ package \cite{Mumax} is employed to numerically solve the LLG equation \eqref{LLG}.

Let us first study the response of the magnetic system as we tune the magnitude of the current. When the current density $J>-1.4\times 10^{12}~\mathrm{A/m^2}$, the SOT is not strong enough to flip the magnetization and thus the magnetization is parallel to the external field ($\mathbf{n}\parallel\mathbf{H}$, left inset of Fig. \ref{simulated_antimagnon}\textcolor{blue}{(a)}). To study the magnon spectrum, we apply a driving $h(t)=h_0 \mathrm{sinc}(\omega_0 t)e_x$ on the whole sample, where $\omega_0/2\pi = 100$ GHz and $h_0 = 50$ mT. The response of spins is recorded and then transformed to Fourier space $(\omega,k_x)$ as
$n_x(k_x,y_0, \omega) = \int n_x(x,y_0,t) e^{-i(k_xx-\omega t)}dx$. The left panel of Fig. \ref{simulated_antimagnon}\textcolor{blue}{(b)} shows a normal parabolic dispersion of magnons, consistent with the theoretical prediction (white line). Besides the uniform mode ($k_y=0$), we notice a series of standing wave modes in the transverse direction ($k_y w=n \pi$, $n=1,2$,...). As the width of the magnetic film increases, the gap of these standing wave mode tends to decrease and they merge with the uniform precession mode ($k_y=0$). When the current density $J<-1.4\times 10^{12}~\mathrm{A/m^2}$ of the SOT is sufficiently strong to switch the magnetization and the antiparallel state ($\mathbf{n}\parallel-\mathbf{H}$) becomes dynamically stable, even though it is not favoured by the Zeeman field. The applied electric current acts as an energy source to maintain the system in this steady state. The excitation spectrum above this dynamically stable state is shown in the right panel of Fig. \ref{simulated_antimagnon}(b). The downward parabola dispersion as predicted by the theory suggests the existence of antimagnons.

\subsection{Vavilov-Cherenkov effect}

The emission of radiation by a charged source moving faster than the speed of sound is called Vavilov-Cherenkov radiation~\cite{cherenkov1934visible,tamm1937coherent,vavilov1934possible}.
This type of radiation applies to more systems than charged particles only.
For instance, the wake generated by a swimming duck may be viewed as an Cherenkov process~\cite{carusotto2013cerenkov}.
Furthermore, it has been shown by Xia~\textit{et al.}~\cite{xia_spin-cherenkov_2016} and Yan~\textit{et al.}~\cite{yan_spin-cherenkov_2013} that the Cherenkov effect can be induced in magnetic systems by moving magnetic field pulses.
One may generalize the definition of Cherenkov radiation even further by noting that the dispersion of the waves has negative energy modes in the rest frame of the uniformly moving source.
Proposals for magnonic Cherenkov radiation using this approach in static magnetic systems using spin-transfer-torques have been considered by de Kruijf~\cite{de_kruijf_spin_nodate} and Koskamp~\cite{koskamp_uence_nodate}.
In these proposals the Cherenkov radiation is induced by a localized transverse field that is static, while the spin transfer torques yield the necessary negative energy modes.  As a result, we expect Cherenkov radiation in the model discussed in the previous section for the case that $ \mathbf{n}_0=-\hat{z} $.
From this point onward we neglect the shape anisotropy $ \Delta\rightarrow0 $ in~\cref{eq:linearized_LLG_equation}, assuming that the qualitative physics still remains intact.

\begin{figure}
	\includegraphics[width=0.9\columnwidth]{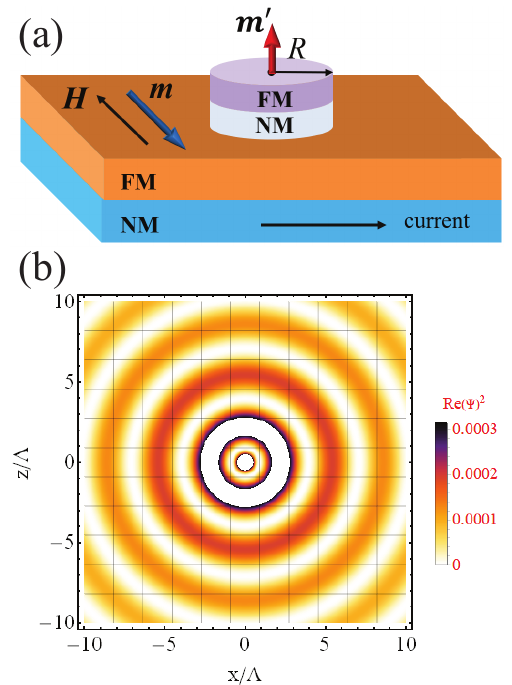}
	\caption{(a) Schematic of the magnetic system to study Cherenkov radiation of spin waves. (b) Density plot of the wave pattern of $ \mathrm{Re}(\Psi)^2 $ in real space for a point like source with $ S_0=1,~h=1 $ and $ I_s=-0.1$.}
	\label{fig:cherenkov-pattern-pointlike-source}
\end{figure}

In the following we investigate the wake pattern associated with the Cherenkov zero modes in a magnetic thin film due to a local transverse magnetic field. This transverse field could be due to a small magnetic pillar on top of the thin film, cf. the proposals of Refs.~[\onlinecite{de_kruijf_spin_nodate}] and~[\onlinecite{koskamp_uence_nodate}].
The set-up we consider is depicted in~\cref{fig:cherenkov-pattern-pointlike-source}\textcolor{blue}{(a)}.
In the presence of an transverse magnetic field, the linearized LLG equation~\eqref{eq:linearized_LLG_equation} becomes
\begin{equation}\label{eq:linearised_LLG_equation-Cherenkov}
	\begin{aligned}
		&\frac{\mathrm{i}(1+\mathrm{i}\alpha)\partial_t}{\gamma\mu_0M_s}\Psi
		+
		\left(h+\Lambda^2\nabla^2-\mathrm{i}I_s\right)\Psi
		=
		S(\mathbf{x}),
	\end{aligned}
\end{equation}
where the localized source $ S(\mathbf{x}) $ is defined in terms of the localized transverse field $ \mathbf{h}(\mathbf{x}) $ via $ S(\mathbf{x})=\mathbf{h}(\mathbf{x})\cdot(\hat{x}+\mathrm{i}\hat{y})/\gamma\mu_0M_s $.
If the local transverse field $ \mathbf{h}(\mathbf{x}) $ is small it may be treated as a source for magnons.
There are several ways to proceed from this point onwards. Here, we will use the Green's function formalism to compute the wake pattern.
For notational convenience we rescale to dimensionless time and frequency $ t\rightarrow t/\gamma\mu_0M_s $ and $ \omega\rightarrow\gamma\mu_0M_s\omega $ and to dimensionless wavevector en length $ \Lambda k\rightarrow k $ and $ x\rightarrow x/\Lambda $.
The Green's function for~\cref{eq:linearised_LLG_equation-Cherenkov} is defined by
$
	\left[
	{\mathrm{i}(1-\mathrm{i}\alpha)\partial_t}
	+
	h+\nabla^2-\mathrm{i}I_s
	\right]
	G(t-t',\mathbf{x-x'})
	=
	\delta(t-t')\delta^2(\mathbf{x-x'}).
$
Since the dispersion relation~\eqref{eq:dispersion-relation-meta-stable-equilibrium} is rotationally invariant in the plane, we will work in cylindrical coordinates $ (\rho\phi,z) $.
In cylindrical coordinates the Green's function equation becomes
\begin{align}
	&\left[
	{\mathrm{i}(1+\mathrm{i}\alpha)\partial_t}
	+
	h-\mathrm{i}I_s
	+\partial^2_\rho+{\rho}^{-1}{\partial_\rho}+{\rho^{-2}}{\partial^2_\phi}
	\right]\\\nonumber
	&G(t-t',\phi-\phi',\rho,\rho')
	=
	{\rho}^{-1}\delta(t-t')\delta(\phi-\phi')\delta(\rho-\rho').
\end{align}
Next we write the Green's function in the form of the following expansion
\begin{align}
	G(\tau,\varphi,\rho,\rho')
	=
	\int \frac{\mathrm d\omega}{2\pi} e^{-\mathrm{i}\omega\tau}\sum_{m}\frac{e^{\mathrm{i}m\varphi}}{2\pi}g_{m}(\omega,\rho,\rho').
\end{align}
Accordingly, the equation of motion for $ g_m(\rho,\rho') $ becomes
\begin{align}\label{eq:Greens-function-equation-cylindrical-coordinates}
	&\left[
		\partial^2_\rho+\frac{\partial_\rho}{\rho}
		+\kappa^2-\frac{m^2}{\rho^2}
	\right]
	g_{m}(\omega,\rho,\rho')
	=\frac{1}{\rho}\delta(\rho-\rho'),
\end{align}
with $ \kappa^2=(1-\mathrm{i}\alpha)\omega+h-\mathrm{i}I_s $.
For $ \rho\neq\rho' $ this is just the equation for the (modified) Bessel functions~\cite[Section 3]{jackson_classical_1998} $ J_n(\kappa \rho) $ and $ H^{(1)}_n(\kappa\rho) $.
In~\cref{app:magnon-greens-function-cylindrical-coordinates} we find that
$
	g_m(\omega,\rho,\rho')
	=
	-({\mathrm{i}\pi}/{2})
	J_m(\kappa\rho_<)H^{(1)}_m(\kappa\rho_>),
$
with $ \rho_<=\min(\rho,\rho') $ and $ \rho_>=\max(\rho,\rho') $,
solves~\cref{eq:Greens-function-equation-cylindrical-coordinates} for the appropriate boundary conditions.
Hence, the magnonic Green's function in cylindrical coordinates and frequency space becomes
\begin{align}\label{eq:greens-function-magon-cylindrical-coordinates}
	&G(\omega,\varphi,\rho,\rho')
	=
	-\frac{\mathrm{i}}{8\pi}
	\sum_{m}{e^{\mathrm{i}m\varphi}}
	J_m(\kappa\rho_<)H^{(1)}_m(\kappa\rho_>).
\end{align}
For simplicity we consider $ \rho>\rho' $, such that $ \rho_<=\rho' $ and $ \rho_>=\rho $.
The wavefunction $ \Psi $ in the presence of the source is accordingly given by
\begin{align}
	\Psi(R,\phi,t)
	=&
	\int \mathrm{d}\omega\mathrm{d}t'\mathrm{d}\phi'\mathrm{d}\rho'
	\\\nonumber&
	e^{-\mathrm{i}\omega(t-t')}G(\omega,\phi-\phi',\rho,\rho')
	S(\rho',\phi',t').
\end{align}
If the source is static and rotationally invariant, the above equation simplifies further to
\begin{align}\label{eq:field-psi-static-source-rotationally-invariant}
	\Psi(\rho)
	=&
	2\pi\int \mathrm{d}\phi'\mathrm{d}\rho'\rho'
	G(0,\phi-\phi',\rho,\rho')
	S(\rho')
	\\\nonumber
	=&
	-\frac{\mathrm{i}\pi}{2}H^{(1)}_0(\kappa \rho)
	\int \mathrm{d}\rho'\rho'
	J_0(\kappa\rho')
	S(\rho'),
\end{align}
We make the following observations.
Far away from a static source, the wavefunction $ \Psi $ behaves proportional to the asymptotic Hankel function $ H_m^{(1)}(\kappa\rho) $ -- higher order Hankel functions may be excited if the sources is not rotationally invariant.
Thus the wavefunction $ \Psi $ is proportional to~\cref{eq:asymptotic-expansion-Bessel-large-rho} for sufficiently large distances $ \rho\gg1/\sqrt{h} $ from the source at the origin, and we find
\begin{equation}\label{eq:asymptotic-behaviour-psi}
	\Psi(\rho)
	\propto
	\sqrt{\frac{1}{\rho}}\exp\left(\mathrm{i}\kappa\rho\right).
\end{equation}
We see that the Cherenkov wave pattern caused by a localized source falls of as $ \exp(I_s\rho/2\sqrt{h})/\sqrt{\rho} $ and oscillates with a wavelength of $ 2\pi/\sqrt{h} $.

In order to make these ideas a bit more concrete, let us consider a point source located at the origin $ S=S_0\delta(\rho')/2\pi\rho' $.
In this case we find that~\cref{eq:field-psi-static-source-rotationally-invariant} becomes
$
	\Psi(\rho)
	=
	-({\mathrm{i} S_0}/{4})J_0(0)H^{(1)}_0(\kappa \rho),
$
with $ J_0(0)=1 $. Using the asymptotic expansion for $ H^{(1)}_0 $, we find that the field sufficiently far from the point source is given by
\begin{equation}\label{eq:field-psi-final-point-source-cylindrical-greens-function}
	\Psi(R)
	\simeq
	-\frac{\mathrm{i}S_0}{4\pi}
	\sqrt{\frac{2\pi}{\sqrt{h}\rho}}
	\exp\left(\frac{I_sR}{2\sqrt{h}}+\mathrm{i}\sqrt{h}\rho-\frac{\mathrm{i}\pi}{4}\right).
\end{equation}
We plot the real part of the Cherenkov wave pattern having for a point-like source with $ S_0=1 $ in~\cref{fig:cherenkov-pattern-pointlike-source}\textcolor{blue}{(b)}. Note that from the wave function $\Psi$ the transverse deviation of the magnetization can be straightforwardly determined. In principle, the wake pattern could be observed experimentally, albeit that it requires high spatial resolution since the wave length of the pattern is of the order of the exchange length, at least for the model that we consider here.

To more clearly bring out the origin of the wake pattern we use an approximate approach which uses the Fourier ansatz for the Green's function around the zero modes in the Green's function.
The existence of these zero modes in the rest frame of the source is a necessary condition for the existence of Cherenkov radiation.
We will see that these zero modes are dominating the Cherenkov wake pattern.
We find the Fourier transformed Green's function becomes
\begin{align}\label{eq:greens-function-fourier-transform}
	(2\pi)^{3}
	G(\omega,\mathbf{k})
	=
	\left[{(1+\mathrm{i}\alpha)\omega}+h-\mathbf{k}^2-\mathrm{i}I_s\right]^{-1}.
\end{align}
If we are just interested in the wake pattern we may suppose that the source is there at all time and thus time independent. We then find
\begin{align}\label{eq:time-independent-field-greens-function-expression}
	\Psi(\mathbf{x},t)
	=&2\pi\int \mathrm{d}\mathbf{x'}\mathrm{d}\mathbf{k}\;
	G(0,\mathbf{k})
	e^{\mathrm{i}\mathbf{k\cdot(x-x')}}
	S(\mathbf{x'})
	\\\nonumber
	=&2\pi\int \mathrm{d}\mathbf{k}\;
	G(0,\mathbf{k})S(\mathbf{k})
	e^{\mathrm{i}\mathbf{k\cdot x}},
\end{align}
where $ S(\mathbf{k}) $ is the Fourier transform of $ S(\mathbf{x}') $.
We continue by defining the one dimensional surface $ \Sigma $ as the points for which $ \mathrm{Re}[G^{-1}(0,\mathbf{k})]=0 $.
Around a point $ \mathbf{k}_0\in\Sigma $ we perform a change of coordinates, for every $ \mathbf k $ in a neighbourhood of $ k_0 $ let $ k_\perp $ be the distance to the surface and $ \mathbf{k}_\parallel(\theta) $ the associated point on the surface~\cite{carusotto2013cerenkov}.
Next we find the field $ \Psi $ to be approximately given by
\begin{align}
	\Psi(\mathbf{x})
	\simeq
	-\frac{S(\mathbf{k}_0)}{(2\pi)^2}\sqrt{h}
	\int_\Sigma\mathrm{d}\theta e^{\mathrm{i}\mathbf{k}(\theta)\cdot \mathbf{x}}
	\int \mathrm{d}k_\perp \frac{e^{\mathrm{i}k_\perp \mathbf{\hat{n}}\cdot\mathbf{x}}}{v_gk_\perp+\mathrm{i}I_s},
\end{align}
with $ \mathbf{v}_g=v_g\mathbf{\hat{n}}=\nabla_\mathbf{k}(\mathbf{k}^2-h)=2\mathbf{k} $ and $ \mathbf{\hat{n}} $ normal to the surface, hence in the above we find $ v_g=2\sqrt{h} $.
We remember that the stability requirement in~\cref{eq:stabilty-sot} gives $ I_s<0 $, which allows us to perform the second integral by closing the contour. This gives
\begin{align}
	\Psi(\mathbf{x})
	\simeq
	-\frac{\mathrm{i}S(\mathbf{k}_0)}{2\pi}\frac{\sqrt{h}}{v_g}
	e^{\frac{I_s\mathbf{\hat{n}}\cdot\mathbf{x}}{v_g}}\Theta(\mathbf{\hat{n}\cdot\mathbf{x}})
	\int_\Sigma\mathrm{d}\theta e^{\mathrm{i}\mathbf{k}(\theta)\cdot \mathbf{x}},
\end{align}
with $ \Theta $ the Heaviside step function.
We determine the final integral via the the stationary phase approximation in which the dominant contribution is assumed to be given by the stationary phase, i.e. $ \partial_\theta(\mathbf{k}(\theta)\cdot \mathbf{x})=0 $.
Using cylindrical coordinates $ (\rho,\phi) $, the above condition becomes $ \sin(\theta-\phi)=0 $, i.e. $ \theta-\phi=\pi/2\pm\pi/2 $.
Since the Heaviside step function only allows $ \mathbf{\hat{n}}\cdot\mathbf{x}>0 $, we are only left with the stationary point $ \theta=\phi $.
By expanding
$ \mathbf{k}(\theta)\cdot\mathbf{x}\simeq\sqrt{h}\rho
-\sqrt{h}\rho(\theta-\phi)^2 /2
$
up to second order in theta around the stationary point and performing the Gaussian integral we end up with
\begin{align}
	\Psi(\rho,\phi)
	\simeq&
	-
	\frac{\mathrm{i}S(\mathbf{k}_s(\phi))}{4\pi}
	\sqrt{\frac{2\pi}{\sqrt{h}\rho}}
	\\\nonumber\times&
	\exp\left(\frac{I_s\rho}{2\sqrt{h}}+\mathrm{i}\sqrt{h}\rho-\frac{\mathrm{i}\pi}{4}\right),
\end{align}
with $ \mathbf{k}_s=\sqrt{h}(\cos(\phi),\sin(\phi)) $ and where we used $ \sqrt{\mathrm{i}}\rightarrow\exp{-\mathrm{i}\pi/4} $.
Hence, we find a static wave pattern that falls off as $ \exp({I_s\rho}/2\sqrt{h})/\sqrt{\rho} $ with an oscillating behaviour of wavelength $ 2\pi/\sqrt{h} $ which is consistent with~\cref{eq:asymptotic-behaviour-psi,eq:field-psi-final-point-source-cylindrical-greens-function}.
\subsection{Thermal occupation of antimagnons}
In this section, we derive the occupation of antimagnons when thermal effects are taken into account and compare it with its magnon counterpart. Following the Green's function formalism developed by Zheng \textit{et al.} \cite{ZhengPRB2017}, we consider a hybrid $\mathrm{FM}|\mathrm{NM}$ structure as shown in Fig. \ref{magnon_antimagnon_setup}\textcolor{blue}{(c)}. For simplicity, we focus on the uniform mode, such that the Hamiltonian becomes $\mathcal{H}=-\omega_a{a^\dagger a}$ with $\omega_a=H>0$. The retarded and advanced Green function of the system are respectively,
\begin{subequations}
\begin{align}
\mathcal{G}_R(\epsilon) &= \frac{1}{\epsilon+\omega_a + i\alpha \epsilon +i \alpha_p(\epsilon-\Delta \mu)},\\
\mathcal{G}_A(\epsilon) &= \frac{1}{\epsilon+\omega_a - i\alpha \epsilon -i \alpha_p(\epsilon-\Delta \mu)},
\end{align}
\end{subequations}
where $\alpha_p$ is the Gilbert damping of the spins due to the interfacial spin pumping, and $\Delta \mu$ is the spin accumulation at the interface. For stability of the antimagnons we need to have $\Delta \mu < -(\alpha+\alpha_p)\omega_a/\alpha_p\equiv\mu_c$. This latter inequality expresses the stability requirement in terms of spin accumulation rather than in terms of spin current as in Sec.~\ref{sec:formalism}.

The spectral function of the excitations is related to the imaginary part of the retarded Green function and it has
two contributions from the bulk damping and interfacial spin pumping as
\begin{subequations}
\begin{align}
&A_\mathrm{bulk}(\epsilon) = \frac{2\alpha \epsilon} {(\omega_a +\epsilon )^2 + (\alpha \epsilon + \alpha_p (\epsilon-\Delta \mu))^2},\\
&A_\mathrm{int}(\epsilon) = \frac{2\alpha_p (\epsilon-\Delta \mu)} {(\omega_a +\epsilon )^2 + (\alpha \epsilon + \alpha_p (\epsilon-\Delta \mu))^2}.
\end{align}
\end{subequations}
The total antimagnon density is found as \cite{ZhengPRB2017}
\begin{equation}\label{antimagnon_density_Green}
n=\frac{1}{2\pi}\int_{-\infty}^\infty d\epsilon \left [ A_\mathrm{bulk}n_B(\epsilon) + A_\mathrm{int}n_B(\epsilon -\Delta \mu) \right ],
\end{equation}
where $n_B(\epsilon)=1/(e^{\beta \epsilon}-1)$ is the Bose-Einstein distribution, with $\beta=1/k_BT$ the inverse thermal energy.

\begin{figure}
	\centering
	\includegraphics[width=0.45\textwidth]{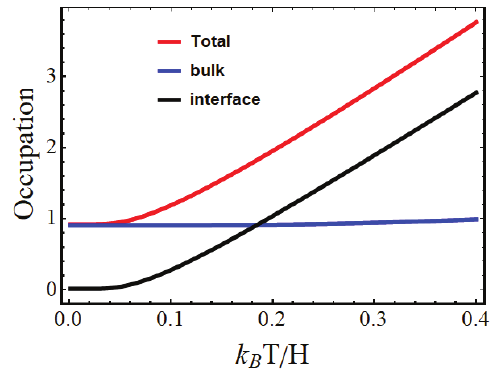}\\
	\caption{Antimagnon density as a function of temperature. Red, blue and black lines represent the total magnon density, bulk contribution and interface contribution, respectively.  The spin accumulation is chosen to guarantee the dynamical stability of the magnetization antiparallel to external field. $\Delta \mu =-1.1 |\mu_c|, \alpha=0.001,~\alpha_p=0.01$.}\label{Green_antimagnon}
\end{figure}

{We assume the dampings to be small $\alpha_p,\alpha \ll 1$, such that we can analytically calculate the magnon density through the technique of counter integration as $n=n_\mathrm{bulk} + n _\mathrm{int}$, where $n_\mathrm{bulk}$ and $n _\mathrm{int}$ respectively refer to the bulk and interface contributions to the antimagnon density that read
\begin{subequations}
	\begin{align}
		&n_\mathrm{bulk} = \frac{1}{e^{-\omega_a/k_BT} -1 } \frac{\alpha \omega_a}{(\alpha + \alpha_p) \omega_a  + \alpha_p \Delta \mu},\\
		&n _\mathrm{int} = \frac{1}{e^{-(\omega_a+\Delta \mu)/k_BT} - 1} \frac{\alpha_p(\omega_a+\Delta \mu)}{(\alpha + \alpha_p) \omega_a  + \alpha_p \Delta \mu}.
		\end{align}
\end{subequations}
Figure \ref{Green_antimagnon} shows that the antimagnon occupation increases with temperature. At low temperature regime, $k_BT \ll \omega_a,\Delta \mu$, the exponential terms in $n_\mathrm{bulk}$ and $n_\mathrm{int}$ are approximately zero, and then the bulk and interface contributions are insensitive to temperature. If we consider a sufficiently thin film, then the interface contribute to the Gilbert damping is sufficiently larger than that from the bulk contribution, i.e. $\alpha_p \gg \alpha$. Therefore the interface contribution to the antimagnon density dominates, as shown in the figure. It is interesting to note the finite occupation of antimagnons even at zero temperature. It may result from the absorption of energy from the reservoir, rather than emission of energy to the reservoir in the equilibrium state \cite{RevModPhys.82.1155}. As the temperature increases, $(e^{-(\omega_a+\Delta \mu)/k_BT} - 1)^{-1} \approx -k_B T/(\omega_a+\Delta \mu)$, such that a linear dependence of antimagnon density on temperature is observed.}

\begin{figure*}
	\centering
	\includegraphics[width=0.85\textwidth]{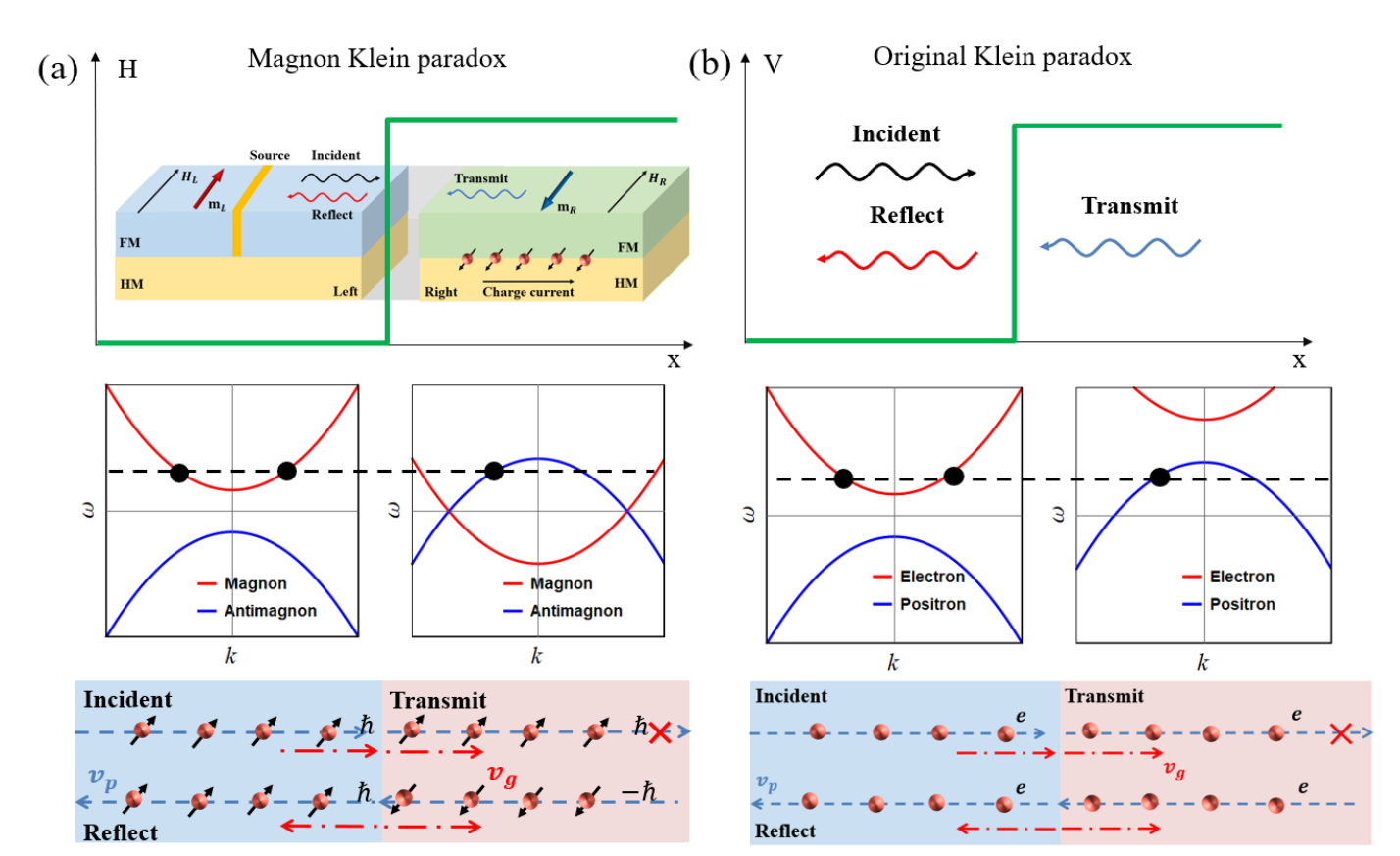}\\
	\caption{Schematic comparison between the original and magnon Klein paradox. The top panel sketches the setup and ``potential" distribution, the middle panel sketches the dispersion relation of particles (red lines) and antiparticles (blue lines), and the bottom panel sketches the flow of incident, reflected and transmitted (spin) currents.}\label{scheme_magnon_Klein}
\end{figure*}

\section{Applications} \label{sec:apps}
In this section we briefly discuss some applications that arise from the coupling between antimagnons and magnons. These are the enhancement of magnon current, engineering of black-hole-horizon analogues for magnons, and magnonic black-hole lasers. For more details we refer the reader to our previous work on these topics in, respectively, Ref.~[\onlinecite{Joren2021}], Ref.~[\onlinecite{roldan2017magnonic}], and Ref.~[\onlinecite{PhysRevLett.122.037203}].
\subsection{Klein effect}
The interaction between magnons and antimagnons allows us to enhance magnon spin currents in hybrid magnetic structures, such as the one shown in Fig. \ref{scheme_magnon_Klein}(a). Here the two magnetic films (left and right) are coupled by and exchange-type interaction, for example, by the RKKY interaction. The magnetization is taken to be parallel and antiparallel to the external fields on the left and right domain, respectively. The external field is carefully designed such that the magnon branch of the dispersion in the left domain overlaps with the antimagnon branch in the right domain. By applying a microwave source in the left domain (orange bar), magnons that travel towards the interface between the left and right magnet are excited. We assuming that the microwave source is such that it excites magnons with frequency in the the frequency range where the magnon and antimagnon dispersions in the left and right magnet, respectively, overlap. When magnon currents with angular momentum $\hbar$ that are excited in the left domain (black dashed line) reach the interface, they will excite backward flowing antimagnon currents ($k_x<0$) with angular momentum $-\hbar$ in the right domain. According to angular momentum conservation, the reflected spin current with angular momentum $\hbar$ is enhanced. Note that a
forward moving flowing current ($k_x>0$) with angular momentum $\hbar$ cannot be excited in the right domain due to the antiparallel orientation of the magnetization.

We dub this magnon current enhancement process as the magnon Klein effect or magnon Klein paradox \cite{Joren2021}. In the original Klein paradox, one considers the scattering of electrons off a potential in the relativistic regime. When the electric potential step is sufficient large such that the positron band in the right region overlaps with the electron band in the incident region, the incoming electrons will excite electron-positron pairs at the interface. As a result, the back-moving electrons enhance the reflection of electrons as shown in Fig. \ref{scheme_magnon_Klein}(b). The energy source of this amplification comes from the energy to maintain the potential landscape. With these similarities, we would like to mention the difference between the magnonic and electronic system. The original Klein effects of electrons are purely relativistic effects, and do not occur for non-relativistic electrons described by the Schr\"odinger equation. The magnonic Klein effects are governed by the linearized LLG equation, which, as we have shown, may be recast as a Schr\"{o}dinger-like equation. The antimagnon occurs when we consider the negative-energy excitations above a metastable state and the Klein effects result from coupling of magnons with these negative-energy excitations. Similar Klein effects may be realized in  analogue gravity systems \cite{Faccio2013}.

\subsection{Antimagnons due to spin-transfer torque}
In this section, we show that antimagnons can also be produced by spin-transfer torques that occur in ferromagnetic metals.
Let us start from the LLG equation with spin-transfer torque~\cite{TataraPRL2004,ZhangPRL2004,ThiavilleEPL2005,ralph2008spin}
\begin{equation}\label{sttllg}
		\frac{\partial \mathbf{n}}{\partial t}=-\gamma \mathbf{n} \times \mathbf{H}_\mathrm{eff} + \alpha \mathbf{n} \times \frac{\partial \mathbf{n}}{\partial t}  +\tau_\mathrm{STT},
\end{equation}
where $\tau_\mathrm{STT}=- (\mathbf{u}\cdot \nabla) \mathbf{n}  + \beta \mathbf{n} \times (\mathbf{u}\cdot \nabla) \mathbf{n}$ is the spin transfer torque that includes the contribution due to spin-conserving and spin-non-conserving interactions between spins of conduction electrons and the magnetic order. Here, the velocity $\mathbf{u}$ is proportional to the electric current density and its polarization. The dimensionless parameter $\beta$ quantifies the degree of spin-non-conserving interactions and is of the same order as the Gilbert damping. Below, we take the velocity $\mathbf{u}$ to be in the $x$-direction.

\begin{figure}
	\centering
	\includegraphics[width=0.49\textwidth]{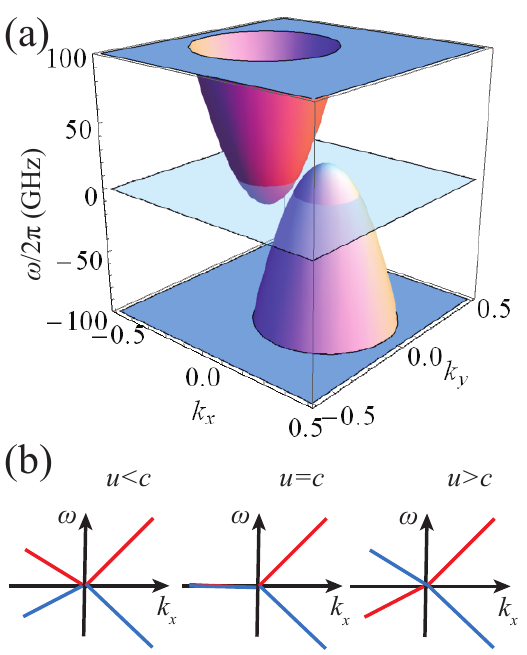}\\
	\caption{{(a) Dispersion relations of magnons and antimagnons in the STT driven case. Parameters of Colbat are used with $A=1.5\times 10^{-11}~ \mathrm{J/m}, M_s = 5.8 \times 10^5~\mathrm{A/m}, u=2.8~\mathrm{km/s}, H=0.1~ \mathrm{T}, \alpha=0.001, \beta=0.0015$.(b) Schematic dispersion of low-energy excitation in an easy-plane magnet subject to an electric current and a small external magnetic field.}}\label{stt_spectrum}
\end{figure}

The classical ground state of the system in the absence of the electric current reads $\mathbf{n}_0=e_z$. To study the excitation above this ground state, {we do the expansion $\mathbf{n} = \mathbf{n}_0 +(\delta n_1 \hat{x}+\delta n_2 \hat{y})$ and linearize the dynamic equations \eqref{sttllg}. Following the similar procedures as that in Sec. \ref{sec:formalism}, we obtain the dispersion relations of the excitations as
	\begin{equation}
		\omega_\pm = -\frac{i\alpha (\Delta + h + \Lambda k^2)}{1+\alpha^2} + \frac{k_xu(1+\alpha \beta) \pm\sqrt{\Sigma} }{1+\alpha^2},
	\end{equation}
	where $\Sigma =( \Delta + h + \Lambda k^2-ik_xu(\alpha-\beta))^2-\Delta^2$. Considering $\alpha, \beta \ll 1$,
	we expand the square root up to the linear order in $\alpha$ and $\beta$ as
	\begin{equation} \label{eq:dispersionstt}
		\begin{aligned}
			\omega_\pm &= -i\alpha (\Delta + h + \Lambda k^2) \pm i k_xu(\beta-\alpha) \\
			&+k_x u\pm \sqrt{(h +\Lambda k^2 )(h+2\Delta + \Lambda k^2) }.
		\end{aligned}
\end{equation}}
The spectrum of the system is found in Fig. \ref{stt_spectrum}\textcolor{blue}{(a)}, where the upward parabola ($\omega_+$) and downward parabola ($\omega_-$) correspond to the magnon and antimagnon excitations, respectively. The system is dynamically stable when $\beta=\alpha$ so that the excitations do not grow exponentially. Due to the spin transfer torques the magnons acquire a Doppler shift $k_x u$ that may be used to engineer magnonic analogues of black-hole-horizons \cite{harms2021dynamically} as we discuss now.


\subsection{Black hole horizons}
We consider now the dispersion relation in Eq.~(\ref{eq:dispersionstt}) for the case that the spin-wave dispersion is linear with velocity $c$. Physically, this could correspond to the situation of an easy-plane magnet at small external field. Ignoring damping, the dispersions are then given by
\begin{equation}
\omega_\pm = \pm c k + u k_x.
\end{equation}
This dispersion is shown in~\cref{stt_spectrum}\textcolor{blue}{(b)}, where red denotes a positive-norm mode and blue denotes a negative-norm mode.
For $u < c$, the spin waves are able to move in both directions. For $u>c$, however, the spin waves are able to travel in one direction only. A magnonic black-hole horizon may be engineering by a spatial transition region between $u<c$ and $u>c$, for example by making the current-carrying wire narrower to increase the current density. Such a transition region then corresponds to a point-of-no-return for spin waves travelling from the $u<c$ to the $u>c$ part.
The same transition region corresponds to a co-called white-hole horizon for spin waves travelling in the opposite direction. The magnon excitations on one side of such horizons overlap in frequency with the antimagnon excitations on the other side, resulting in magnon-antimagnon coupling across the horizons. In the quantum regime, this coupling yields Hawking radiation of spin waves away from the horizon~\cite{roldan2017magnonic}.
Furthermore, one may view the set-up in~\cref{scheme_magnon_Klein}\textcolor{blue}{(a)} as a generalized black-hole horizon.
In such set-ups emission of magnon-antimagnon pairs also occurs.
But the fact that these magnons are dispersive causes the spectrum of the pair production to be non-thermal.

\subsection{Lasing}
We have argued that coupling of spatially separated magnons to antimagnons may lead to enhanced reflection and stimulated emission such as Hawking radiation.
When the region containing antimagnons has a finite size, the antimagnons can acquire closed orbits of specific frequencies.
The latter results in a dynamical instability with self amplification of the bounded modes~\cite{corley1999black,coutant2010black} analogous to mode coalescence in exceptional points~\cite{coutant2016dynamical}.
In Ref.~[\onlinecite{PhysRevLett.122.037203}] we considered such a set-up in which a confined region with antimagnons between a black-hole and white-hole horizon is considered.
The confined region gives rise to a finite set of self amplified modes of which the frequencies can approximately be obtained via Born-Sommerfeld quantization.
The onset of the instability is signalled by a large increase in the scattering of magnons off the confining region, leading to large amplification of spin waves that are incident on the confining region.
More concretely, the scattering amplitudes near the complex eigenfrequencies of self amplified modes $ \lambda_a=\omega_a+\mathrm{i}\Gamma_a $ behave like Lorentzians $ \sim|\omega-\omega_a-\mathrm{i}\Gamma_a|^{-2} $, which are sharply peaked around the self amplified resonance frequencies.

\section{Conclusions and outlook}
In this perspective, we have discussed how to stabilize negative-energy excitations of magnetic systems which we refer to as antimagnons. We have discussed several set-ups in which antimagnons interact with conventional magnons. One the one hand, this leads to novel schemes for spin-wave amplification. On the other hand, these set-ups provide platforms to study analogues of high-energy and gravitional physics in magnetic systems.

For future work, it would be interesting to consider the non-linear regime in more detail. Moreover, it would be interesting to generalize the ideas presented here to different types of magnetic order and different ways of driving the system to stabilize the negative-energy modes.

We hope that this perspective motivates experiments to utilize antimagnons in both fundamental and applied physics.

\section{Acknowledgments}

H.Y.Y acknowledges the European Union's Horizon 2020 research and innovation programme under the Marie Sk{\l}odowska-Curie Grant Agreement SPINCAT No. 101018193. R.A.D. is member of the D-ITP consortium that is funded by the Dutch Ministry of Education, Culture and Science (OCW). R.A.D. has received funding from the European Research Council (ERC) under the European Union's Horizon 2020 research and innovation programme (Grant No. 725509). This work is in
part funded by the projects “Black holes on a chip” with
project number OCENW.KLEIN.502 and by the Fluid Spintronics research programme with project number 182.069, which
are financed by the Dutch Research Council (NWO).

\section{Data availability}
Data sharing is not applicable to this article as no new data were
created or analyzed in this study.

\section*{References}
\bibliography{./antimagnonics}
\appendix
\section{Magnon Green's function in cylindrical coordinates}\label{app:magnon-greens-function-cylindrical-coordinates}
In this section we determine the Magnon Green's function in cylindrical coordinates.
The Green's function of~\cref{eq:linearised_LLG_equation-Cherenkov} in cylindrical coordinates is given by
\begin{align}
	&\left[
	{\mathrm{i}(1-\mathrm{i}\alpha)\partial_t}
	+
	h-\mathrm{i}I_s
	+\partial^2_\rho+{\rho}^{-1}{\partial_\rho}+{\rho^{-2}}{\partial^2_\phi}
	\right]\\\nonumber
	&G(t-t',\phi-\phi',\rho,\rho')
	=
	{\rho}^{-1}\delta(t-t')\delta(\phi-\phi')\delta(\rho-\rho').
\end{align}
We may write down the above Green's function in the form of a product expansion by separating the variables
\begin{align}\label{eq:magnon-greens-function-appendix}
	G(\tau,\varphi,\rho,\rho')
	=
	\int \frac{\mathrm d\omega}{2\pi} e^{-\mathrm{i}\omega\tau}\sum_{m}\frac{e^{\mathrm{i}m\varphi}}{2\pi}g_{m}(\omega,\rho,\rho').
\end{align}
Accordingly, the equation of motion for $ g_m(\rho,\rho') $ becomes
\begin{align}\label{eq:Greens-function-equation-cylindrical-coordinates-appendix}
	&\left[
	\partial^2_\rho+\frac{\partial_\rho}{\rho}
	+\kappa^2-\frac{m^2}{\rho^2}
	\right]
	g_{m}(\kappa,\rho,\rho')
	=\frac{1}{\rho}\delta(\rho-\rho'),
\end{align}
with $ \kappa^2=(1-\mathrm{i}\alpha)\omega+h-\mathrm{i}I_s $.
For $ \rho\neq\rho' $ this is just the equation of the (modified) Bessel functions~\cite[Section 3]{jackson_classical_1998} $ J_n(\kappa \rho) $ and $ H_n(\kappa\rho) $.
In principle the solution of the above equation of motion can be written as a linear superposition of $ I_n $ and $ H^{(1)}_n $,
the asymptotic behaviour of $ J_m$ and $ H^{(1)}_m $ for $ \kappa\rho\ll1 $ is described by
\begin{subequations}\label{eq:asymptotic-expansion-Bessel-small-rho}
	\begin{align}
		\kappa\rho\ll1,
		~~~
		J_m(\kappa\rho)&\rightarrow\frac{1}{\Gamma(m+1)}\left(\frac{\kappa\rho}{2}\right)^m,
		\\
		H^{(1)}_m(\kappa\rho)&\rightarrow
		\left\{
		\begin{array}{r r}
			\frac{2\mathrm{i}}{\pi}\ln\left(\frac{\kappa\rho}{2}\right)~~\;,&m=0,\\[3pt]
			-\frac{\mathrm{i}\Gamma(m)}{\pi}\left(\frac{2}{\kappa\rho}\right)^m,&m\neq0,
		\end{array}\right.
\end{align}
\end{subequations}
Moreover, the asymptotic behaviour for $ \kappa\rho\gg1 $ is described by
\begin{subequations}\label{eq:asymptotic-expansion-Bessel-large-rho}
	\begin{align}
		\kappa\rho\gg1,
		~~~
		J_m(\kappa\rho)&\rightarrow\sqrt{\frac{2}{\pi\kappa\rho}}\cos\left(\kappa\rho-\frac{m\pi}{2}-\frac{\pi}{4}\right),
		\\
		H^{(1)}_m(\kappa\rho)&\rightarrow\sqrt{\frac{2}{\pi \kappa\rho}}\exp\mathrm{i}\left(\kappa\rho-\frac{m\pi}{2}-\frac{\pi}{4}\right),
	\end{align}
\end{subequations}
with $ \Gamma(n)=(n-1)! $ the the gamma function.
We proceed by letting $ \psi_1(\kappa\rho) $ be a linear combination which is regular at $ \rho\rightarrow0 $, satisfying the correct boundary condition for $ \rho<\rho' $. And furthermore, let $ \psi_2(\kappa\rho) $ be the linear combination of Bessel functions such that it is regular at $ \rho\rightarrow\infty $, satisfying the boundary conditions for $ \rho>\rho' $.
By remembering that the stability requirement in~\cref{eq:stabilty-sot} states that $ I_s<0 $, we find the above boundary conditions are satisfied if $ \psi_1(\kappa\rho)=A_mJ_m(\kappa\rho) $ and $ \psi_2(\kappa\rho)=B_mH^{(1)}_m(\kappa\rho) $
The symmetry in $ \rho $ and $ \rho' $ of the Green's function requires
$
g_m(\kappa,\rho,\rho')=\psi_1(\kappa\rho_<)\psi_2(\kappa\rho_>),
$
with $ \rho_<=\min(\rho,\rho') $ and $ \rho_>=\max(\rho,\rho') $.
At this point we still need to fix the constant $ A_mB_m $, which is determined by the delta function in~\cref{eq:Greens-function-equation-cylindrical-coordinates-appendix}.
This in done by integrating~\cref{eq:Greens-function-equation-cylindrical-coordinates-appendix} over an $ \epsilon $ distance around the point $ \rho=\rho' $.
We furthermore consider the boundary conditions in which $ g_m(\kappa,\rho,\rho') $ is continuous everywhere and its derivative is continuous everywhere except at the point where the Dirac delta becomes infinite.
Integration around $ \rho=\rho' $ gives
\begin{equation}
	\lim_{\epsilon-\rightarrow0}
	\left(
	\left.\frac{\mathrm{d}g}{\mathrm{d}\rho}\right\rvert_{\rho'+\epsilon}
	-
	\left.\frac{\mathrm{d}g}{\mathrm{d}\rho}\right\rvert_{\rho'-\epsilon}
	\right)
	=
	\frac{1}{\rho'}.
\end{equation}
This in turn gives
$
	\psi_2\psi_1'-\psi_1\psi_2'={1}/{\kappa\rho'}.
$
Since the above expression should hold anywhere, we determine $ A_mB_m $ using the asymptotic expansions of $ J_m $ and $ H_m^{(1)} $~\cref{eq:asymptotic-expansion-Bessel-small-rho,eq:asymptotic-expansion-Bessel-large-rho}. We find
\begin{equation}
	A_mB_m\frac{2\mathrm{i}}{\pi}\frac{1}{\rho'}=\frac{1}{\rho'},
\end{equation}
and hence $ A_mB_m=-\frac{\mathrm{i}\pi}{2} $.
This yields
\begin{align}
	g_m(\kappa,\rho,\rho')
	=
	-\frac{\mathrm{i}\pi}{2}
	J_m(\kappa\rho_<)H^{(1)}_m(\kappa\rho_>).
\end{align}
The magnon Green's function in~\cref{eq:magnon-greens-function-appendix} thus becomes
\begin{align}
	&G(\tau,\varphi,\rho,\rho')
	=\\\nonumber
	&-\frac{\mathrm{i}}{8\pi}
	\int {\mathrm d\omega} e^{-\mathrm{i}\omega\tau}
	\sum_{m}{e^{\mathrm{i}m\varphi}}
	J_m(\kappa\rho_<)H^{(1)}_m(\kappa\rho_>).
\end{align}

\end{document}